# Progress and Prospects in Weather and Climate Modelling

R. Krishnan, Manmeet Singh, Ramesh Vellore and Milind Mujumdar

Centre for Climate Change Research
Indian Institute of Tropical Meteorology, Pune 411008

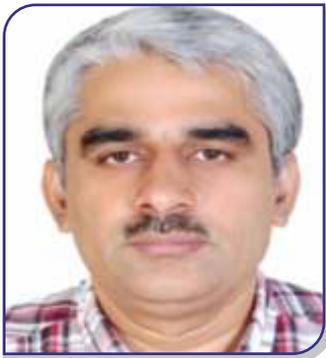

Dr. Krishnan specializes in climate modeling to address scientific issues relating to the "Dynamics and variability of the Asian monsoon, Global climate change and impacts on monsoon precipitation extremes, Phenomenon of monsoon-breaks and droughts". Currently he is leading the Centre for Climate Change Research (CCCR) at the Indian Institute of Tropical Meteorology (IITM), Pune as Executive Director and is involved in developing in-house capability in Earth System Modeling to address the science of climate change. He is a Fellow of the Indian Academy of Science and Indian Meteorological Society(IMS) of India. He is a member Asia Australia Monsoon panel, CORDEX Science Advisory team and Scientific Steering committee, Monsoon Asia Integrated Regional Study. He is the coordinating lead author sixth assessment report, IPCC WG1. He has been awarded the Prof. A.Vernekar award of the IMS. He is the editor of the Earth System Dynamics, European Geophysical Union.

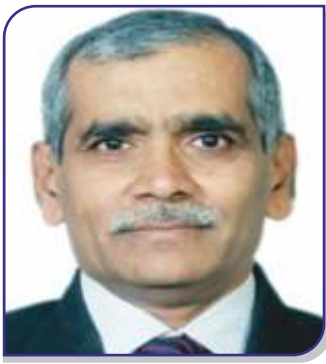

Dr. Milind Mujumdar is a senior scientist at the Centre for Climate Research Centre (CCCR), Indian Institute of Tropical Meteorology (IITM), Pune. His scientific research interests include (1) Asian monsoon system and its response to climate change using observations and climate modelling studies (2) Evaluation and assessment of the regional climate downscaling over South Asia (CORDEX). Currently he is leading a scientific project at IITM for soil moisture measurements using the Cosmic Ray Soil Moisture Observation System (COSMOS), to better quantify the feedbacks between land-surface processes and the Indian monsoon. Dr. Mujumdar is passionately involved in popularizing science education and outreach activities.

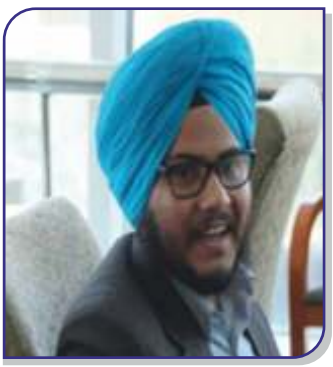

Mr Manmeet Singh is a Scientist at the Centre for Climate Research Centre (CCCR), Indian Institute of Tropical Meteorology (IITM), Pune. He is presently pursuing his PhD in Climate Studies from IIT Bombay. He has an extensive experience in computational modelling of earth system and its processes. He is presently involved in studying the Monsoons using coupled general circulation models and non-linear methods. He also has experience in development of Direct Numerical Simulation model and using it for basic understanding of jets. He is passionate about Deep Learning as a tool to solve the problems in Earth Sciences.

This popular article provides a short summary of the progress and prospects in Weather and Climate Modelling for the benefit of high school and undergraduate college students and early career researchers. Although this is not a comprehensive scientific article, the basic information provided here is intended to introduce students and researchers to the topic of Weather and Climate Modelling - which comes under the broad discipline of Atmospheric / Oceanic/ Climate /Earth Sciences. This article briefly summarizes the historical developments, progress, scientific challenges in weather and climate modelling and career opportunities.

## Introduction to Weather and Climate Models:

The scales of atmospheric motion are closely tied to the spatial and temporal dimensions of the phenomenon (Figure. 1). The spatial scales range from micro (~ 100 m) through planetary scales (~ $10^8$) and the temporal variations from seconds to centuries. Microscale motions (e.g. dust devils, wind gusts) are at the lower end of the spectrum with spatial range < 1 km and persisting under minutes. Mesoscale phenomena with spatial dimensions O (100 km) typically last from several minutes to a few hours (e.g., thunderstorms, tornadoes, land-sea breeze). Tropical clouds can range from individual cloud clusters (scale ~ tens of km) to large-scale organized cloud systems (scale ~ a few thousands of kilometres). Synoptic weather systems, large-scale and planetary scale motions have large spatial scales exceeding thousands of kilometres and last beyond a day up to several days and even a couple of months (e.g., the Madden Julian Oscillation, MJO). Climatic phenomena such as South Asian Monsoon, El Nino and La Nina are associated with global or planetary scale dimensions and can last from seasons to years. Weather and climate models are tools designed to understand and predict the evolution of the diverse scales of motion and also interactions across different scales.

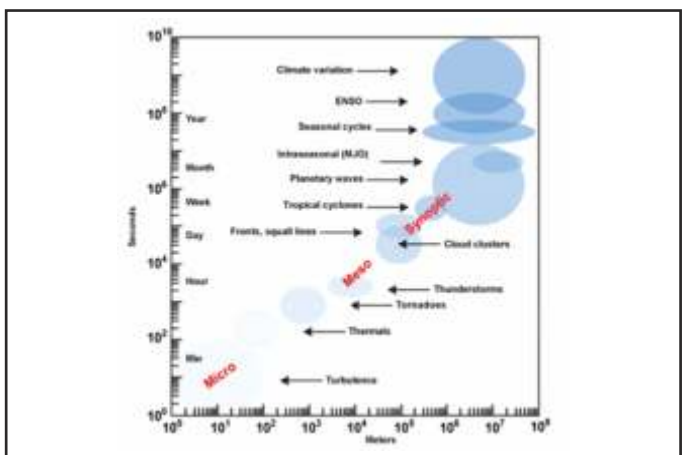

Figure.1: Scales of Atmospheric Motion, adapted from The COMET Program

## History and development of weather and climate models

Stocker (2014) has provided a comprehensive description of the various developments in weather and climate modelling. "Weather is what you see and climate is what you expect" is a famous adage. Basically weather represents short term (minutes to days) variations in the atmospheric parameters viz., temperature, humidity and winds, while climate is a slowly varying feature of the atmospheric system characterized by averages of specific states of the atmosphere.

*Numerical Weather Prediction (NWP):* In 1904, the Norwegian meteorologist Vilhelm Bjerknes (1862-1951) was the first to recognize the prediction of the state of the atmosphere as an initial-value problem of solving a system of mathematical equations (i.e., nonlinear partial differential equations) governing the physical laws of atmospheric dynamics. The first numerical weather prediction (NWP) was attempted by the Englishman Lewis Fry Richardson (1881-1953) through a numerical integration of equations in 1917. Richardson made use of the so-called primitive equations, consisting of conservation equations for momentum, energy and mass and the ideal gas equation, and utilized observations of temperature and pressure from stations across Europe as initial conditions. An attempt of realizing the numerical 6-h forecast of surface pressure change calculated by hand unfortunately turned out to be quite unrealistic, and this novel pursuit was published in "Weather prediction by numerical processes (Richardson 1922)". It was recognized much later, that the Richardson's failed forecasts were due to the fact that atmospheric conditions were not balanced at the initial time of the numerical simulation, and furthermore violations in the numerical discretization were attributed to the requirement of model time step compatibility with model grid or mesh size (Courant et al. 1928). In continuation with this, the Swedish American meteorologist Carl-Gustav Rossby showed that conservation of absolute vorticity (fluid spin

+ planetary spin) in atmospheric motions could provide a large-scale perception of the observed movement of atmospheric centers of action (Rossby 1939). With the advent of first electronic computer ENIAC in 1950, the atmospheric scientists Charney, Fjortfort and von Neumann proposed a simplified system of equations and finally made the first ever encouraging NWP of mid-tropospheric height fields (Charney et al. 1950), which marked the milestone beginning of NWP. Norman Phillips (1955) conducted the first long numerical simulations of a simple atmospheric model, leading to the beginning of the development of general circulation models (GCM), which solved the atmospheric equations of flow. Considerable efforts commenced in late 1950s for the use of multi-level primitive equation models for operational forecasting and many other meterorological services. Owing to hydrostatic balance assumed in the primitive-equation models which cannot deal with convection scale motions explicitly, convective adjustment methods were developed to avoid unstable solutions (Manabe and Strickler 1964) and were followed by more sophisticated representations of the interactions of moisture, convective clouds and environment (Kuo 1965, 1974, Arakawa and Schubert 1974). In addition, representations of sub-grid scale physical processes in terms of the large-scale variables (eg., dry and moist convection, clouds, radiation, boundary layer, turbulence etc) were incorporated into NWP models. Furthermore, advanced mathematical techniques for providing the balanced initial states to these models were also evolved in time (e.g., Baer and Tribbia 1977, Machenhauer 1977; Daley 1991).

In the 1960s, Edward Lorenz performed numerical integrations of a simplified system of nonlinear equations of forced deterministic hydrodynamic flow, for a longer period, which was computationally expensive at that time. In order to save computational time, Lorenz performed a sliced simulation by restarting the simulation at a mid-point with a truncated initial condition, with the expectation that the results would be identical to the initially planned long simulation. However, he was perplexed to notice that the two solutions, starting from two different initial states, were radically different. This laid the foundation for weather predictability, wherein he demonstrated the criticality of initial conditions in weather forecasts. This pioneering work of Lorenz (1963) paved way for understanding the limits of predictability and emphasized that the upper limit on predictability of weather is about two weeks.

Notwithstanding these inherent weather forecasting challenges, NWP has undergone a quiet revolution during the last few decades owing to the steady accumulation of scientific knowledge and technological advances over time (Ref: Bauer et al. 2015). Modern NWP basically involves determining the future atmospheric state by solving a coupled nonlinear system of PDEs involving billions of computations per time step starting from the initial time to weeks and months ahead, which necessitates the use of powerful computers for NWP. Substantial improvements in producing realistic atmospheric initial states have emerged through major developments in Data Assimilation (DA) techniques: The process of ingesting real-time observations from multiple platforms (eg., satellite retrievals, radiosondes, aircraft and ship observations, surface observations over land, ocean and ice-covered areas) with the first guess (3-6 hour model forecasts) (eg., Kalnay 2003; Navon, 2009; Law et al. 2015). The DA cycle is routinely performed on a daily basis at all the weather prediction centres across the globe (e.g., National Centre for Medium Range Weather Forecasting [NCMRWF, Noida, India], European Centre for Medium Range Weather Forecasts [ECMWF, Reading, UK], National Centre for Environmental Prediction [NCEP, USA], Japan Meteorological Agency [JMA, Tokyo, Japan], Bureau of Meteorology [BMRC, Melbourne, Australia], Korean Meteorological Administration [KMA, Seoul,

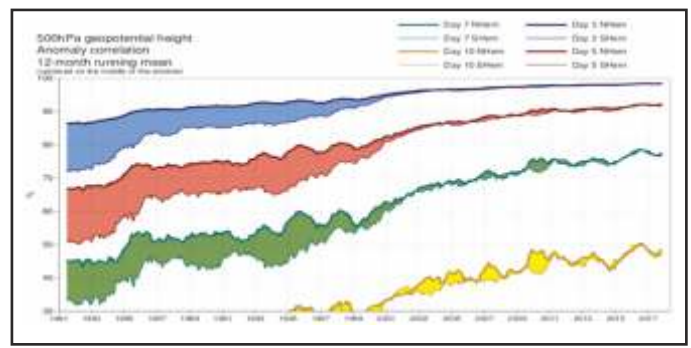

Figure. 2 shows the forecast skill at 3-day, 5-day, 7-day and 10-day ranges computed over the extra-tropical northern & southern hemispheres, based on the ECMWF forecasts [Source: https://www.ecmwf.int/en/forecasts/charts/catalogue/plwww_m_hr_ccaf_adrian_ts?facets=undefined&time=2018061100

South Korea] and others). In summary, major advances in NWP during the last few decades have emerged from (a) Developments of sophisticated data assimilation techniques for improved model initialization (b) Incorporation of enhanced observational coverage both in space and time (c) Vast amount and diversity of satellite observational data (d) Improvements in representing unresolved processes in global models (e)

Advent of ensemble methods for producing forecast uncertainty estimates and (f) Major advancements in high-performance computing (HPC) and increased scalability of NWP model codes (Bauer et al. 2015).

The evolution of the ECMWF model forecast skills is shown for the period 1981-2017 in Figure. 2. The skills are objectively and quantitatively assessed, by comparing the forecast with what actually happens every day. It can be seen that weather forecast skills have significantly increased over the past 40 years. It can also be seen that, forecast skill in the lead time of 3 to 10 has been consistently increasing by about one day per decade, implying that today's 7-day forecast is as accurate as the 5-day forecast twenty years ago (Bauer et al. 2015). In particular, substantial improvements in the predictive skill can be noted in the Southern Hemisphere (which is at present comparable to that of the Northern Hemisphere). This improvement largely comes because of increased coverage of satellite based information used in data assimilation over the ocean dominated Southern Hemisphere. In the Indian context, very high resolution (grid size ~ 27 km) short-range ensemble weather forecasts are being issued on a daily basis (http://srf.tropmet.res.in/srf/# ), which is also being upgraded to finer grid resolutions (grid size 12 km). These high-resolution short-range weather forecasts are proving to be quite promising with regard to high-impact weather systems and heavy precipitation events, particularly during the summer monsoon season.

Satellite data have progressively become the predominant source of information assimilated in NWP models. While the main advantage of satellite data is that they provide a uniform spatial and temporal coverage of the atmosphere, it is equally important to exploit the information contained in such data using advanced data assimilation techniques. Enhancements in satellite observations (i.e., remote sensing sensors) together with improvements in weather and climate modelling and major advances in data assimilation techniques have led to significant improvements in the skill of short and medium-range weather forecasts, particularly over the extra-tropics (https://www.ecmwf.int/en/research/data-assimilation/observations). Lack of observations of atmospheric winds globally has been a major limitation for NWP over the tropics and monsoon regions. The recently developed Aeolus satellite by the European Space Agency (ESA) now makes it possible to measure wind profiles from space using Doppler-wind Lidars (https://www.bbc.com/ news/science-environment-44415752). Exploiting the space-borne wind information is expected to significantly enhance the accuracy of NWP forecasts over the tropics and monsoon regions in the near future.

**Predicting beyond the weather scales:** While short and medium range (time-scale < 10 days) predictability of weather primarily comes from the atmospheric initial conditions, extending the predictions to longer time-scales, such as sub-seasonal / intra-seasonal (> 2 weeks), seasonal, decadal and longer time-scales is dependent on slowly varying boundary conditions like sea-surface temperature (SST), snow-cover, etc. (Charney and Shukla, 1981). The hypothesis proposed by Charney-Shukla remains the central paradigm for seasonal monsoon predictability research for nearly four decades. Climatic phenomena, such as the El Nino / Southern Oscillation (ENSO), Indian Ocean Dipole (IOD), Modoki, exert enormous socio-economic impacts across the globe (http://www.jamstec.go.jp/frcgc/research/d1/iod/e/seasonal/overview.html). Slowly evolving SST boundary conditions on these long-time scales involve ocean-atmosphere interactions, thereby necessitating the use of coupled ocean-atmosphere models for simulating and predicting low-frequency climatic variations beyond the weather scale. Additionally, coupled models are powerful tools to understand and predict the influence of human activities since the 19th century, which have led to rapid increase in the concentration of atmospheric greenhouse gases and aerosols, in altering the climate system – the so-called Climate Change problem.

**Climate Modelling:** Nearly two decades after the development of first atmospheric models, ocean models were conceptualised (Bryan and Cox, 1967). It was also realized that atmospheric and ocean models need to be dynamically coupled to better understand the climate processes, and the first coupled model was developed by Syukuro Manabe and his team in 1960s at the Geophysical Fluid Dynamics Laboratory (GFDL), Princeton, USA (Manabe and Bryan 1969). Significant improvements in ocean-atmosphere coupling have come along in 1990s with the addition of different components of the climate system viz., sea-ice, carbon cycle, land use, vegetation type and biogeochemistry processes of the land and oceanic areas (Stocker et al. 2014 and see the references Atlantic oceans [https://www.pmel.noaa.gov/gtmba/ ], ARGO floats which routinely measure vertical profiles of temperatures, salinity and other parameters in the global oceans [ http://www.argo.ucsd.edu/About_Argo.html ], together with satellite-derived sea-level variations

[https://sealevel.jpl.nasa.gov/missions/topex/], surface-winds [ https://manati.star.nesdis.noaa.gov/datasets/ASCATData.php/], precipitation products [ https://pmm.nasa.gov/content/ ] and several others and developments in ocean and atmosphere data assimilation. All these developments have contributed to major improvements in subseasonal-to-seasonal predictions.

Sub-seasonal-to-Seasonal (S2S) prediction is an active area of research for bridging the prediction gap between weather and climate forecasts (http://s2sprediction.net/). Currently operational weather prediction centres and several research institutions are actively contributing to sub-seasonal forecasting of tropical phenomena like the MJO and the boreal summer monsoon intra-seasonal oscillations. Significant progress has been achieved in recent times to skilfully forecast the MJO and the active / break spells of the boreal summer monsoon in about 2-3 weeks in advance (eg., Joseph et al. 2015, Wheeler et al. 2016, Marshall et al. 2016, Vitart, 2017, http://s2sprediction.net/ ). Considerable improvements in the skill of seasonal forecasts of the Indian monsoon rainfall, using ocean-atmosphere coupled models, are also being realized (Ref: Pillai et al. 2018, http://www.imdpune.gov.in/Clim_Pred_LRF_New/). Several climate prediction centres around the globe routinely issue forecasts of the evolution of ENSO, IOD, Modoki and their associated climatic impacts around the globe on real-time (eg., http://www.jamstec.go.jp/apl/e/ , http://cfs.ncep.noaa.gov/, https://www.ecmwf.int/en/forecasts/charts/seasonal/, https://iri.columbia.edu/our-expertise/climate/ forecasts/seasonal-climate-forecasts/ and others).

Modelling of the climate and earth systems has been evolving as an interdisciplinary science (Figure. 3). During the last few decades, there have been substantial improvements in climate modelling that led to better understanding of our earth system and enhanced capabilities to make climate projections. The role of high performance scientific computing has been crucial in these developments. Climate Change is one of the major threats to the world today owing to its far-reaching implications for environment, agriculture, water availability, natural resources, ecosystems, biodiversity, economy and social well-being. Recent advances in climate modelling enable us to study the climatic response to increasing greenhouse gases, anthropogenic aerosols. There is strong scientific evidence pointing to the role of human activities in altering the Earth's climate through a rapid rise in the concentration of atmospheric greenhouse gases (GHG) since the 19$^{th}$ century (IPCC, 2013). Observations of the climate system based on direct measurements and remote sensing from satellites and other platforms indicate that the warming of the climate system has been unequivocal since 1950s and many of the observed changes are unprecedented over decades to millennia (Stocker et al., 2013). Long-term climate model simulations, that took part in the IPCC AR5 report, provide very high confidence in interpreting the observed global-mean surface temperature trends during the 20$^{th}$ and early 21$^{st}$ centuries; and the human influence on the climate system (Stocker et al., 2013).

Earth System Modelling (ESM) is an advanced modelling approach to comprehend the interactions of atmosphere, ocean, land, ice, and biosphere to estimate the state of regional and global climate under a wide variety of conditions (Figure. 3). ESMs allow us to investigate the nonlinear behavior and multi-scale variability of the climate system, as well as to make predictions of its future states of the climate variables under a wide variety of conditions. In recent times, ESMs are being widely used to understand the impacts of human-induced forcing (eg., enhanced greenhouse gas and aerosol emissions, land use and land cover changes, etc.) on the climate system. ESMs from several climate modelling groups around the world have been contributing to the Coupled Modelling Intercomparison Project (CMIP: Meehl et al. 2007, Taylor et al. 2012), which is coordinated by the World Climate Research Programme (WCRP) and forms the basis for the climate projections in the Intergovernmental Panel on Climate Change (IPCC) Assessment reports. For the first time from India, the IITM-ESM from the Indian Institute of

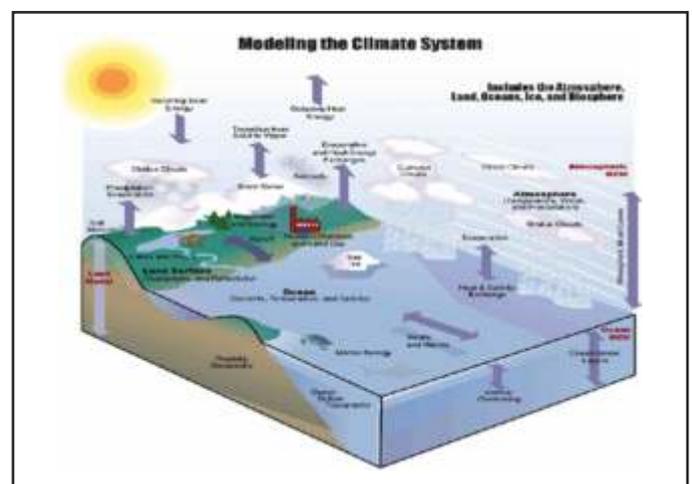

Figure. 3: Schematic of the state-of-the-art climate models. Representing complex interactive feebacks in the climate system necessitates a modular approach in climate model development. (Adapted from https://www.hpcwire.com/2006/07/21/the_next_generation_of_climate_models-1/).

Tropical Meteorology, Pune, will be contributing to the upcoming CMIP Phase-6 experiments and the IPCC Sixth Assessment Report (AR6) (Ref: Swapna et al. 2015, 2018).

## Scientific challenges in tropical weather and climate prediction:

There have been significant advances in weather and climate modelling during the last few decades. Substantial improvements in the skill of short and medium range weather forecasts have been realized, particularly over the extra-tropical regions. However, weather and climate predictability over the tropics and monsoon dominated areas has somewhat progressed at marginal pace over the years due to a variety of challenges posed in modelling the tropical phenomena. Below are some of the key scientific challenges that require significant research and development in the future:

- Unlike extra-tropics, tropical and monsoon circulations are strongly tied to clouds and precipitation processes. TRMM satellite observations and model experiments demonstrate the importance of the 3-dimensional structure of latent heat of condensation in driving tropical and monsoon circulation on different space and time scales (e.g., Houze, 1997, Mapes, 1993, Schumacher et al. 2004, Krishnamurti et al. 2010, Choudhury and Krishnan 2011, Choudhury et al. 2018). Yet, there are large inadequacies in the representation of the interaction between tropical cloud systems and circulation in weather and climate models. This still remains a fundamental research problem, which needs to be addressed in order to particularly improve predictions of monsoon precipitation, tropical cyclones, extreme rainfall events, etc.

- Improving the representation of ocean-atmosphere coupled processes will be crucial to accurately represent the exchange of fluxes of momentum, heat and moisture across the air-sea interface, in order to capture low-frequency climatic variations like the Madden Julian Oscillation, the active and break spells of the Indian monsoon, El Nino / Southern Oscillation (ENSO) in the Pacific and the Indian Ocean Dipole (IOD) in the tropical Indian Ocean, among several others.

- Development of clouds and precipitation involve processes on smaller scales (eg., micro and meso- scales), which are influenced by aerosols (suspended particles in the atmosphere) of different sizes, their concentration and composition and interactions with radiation (Stevens and Feingold, 2009). Understanding the aerosol, radiation and cloud interactive feedbacks and representing them in numerical models is an active area of current research and development (Figure. 4).

- Representation of land-atmosphere, cryosphere-atmosphere interactions (eg., atmospheric feedbacks from soil-moisture and snow variations), climatic variations due to land use and land cover changes are also important areas for scientific research. In particular, a comprehensive understanding of the feedback mechanisms among the different components of the Earth System is essential to quantify changes in the Water Cycle, Snow and Glaciers (eg., Himalayan snow cover) and sea-level rise on global and regional scales, in a changing climate.

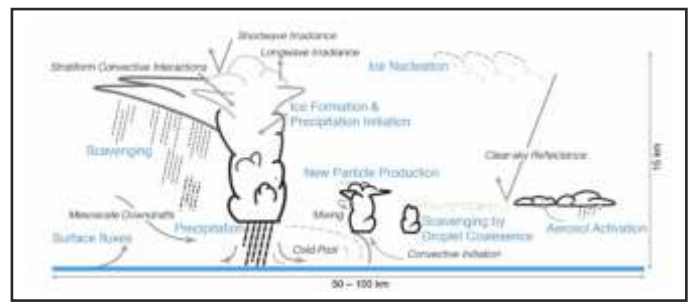

Figure. 4: Aerosol Cloud Interactions (Courtesy: Prof. Robwood, University of Washington).

## Opportunities:

The explosive growth of weather and climate modelling globally during the last decade has created vast opportunities with emergent newer areas (eg., artificial intelligence, nonlinear dynamical systems, complex networks, advanced data analysis techniques, etc.) for a wide range of scientific and socio-economic applications. Technological advances in computing and an ever growing need of big data processing systems also warrant technically skilled workforce to untangle the multi-scale nature of weather and climate processes. Development of capabilities in weather and modelling (including data assimilation) offers a wide range of career opportunities both in India and abroad [eg., National services (eg., Meteorological, Defence, Agriculture, Water Resources, Forestry, Health, Fishery, Renewable Energy, Space and Nuclear Energy, among many others), National and International research laboratories – related to atmospheric, oceanic, climate, earth and environmental science programmes, University Departments, High performance and scientific computing industry, Aviation and shipping industry, Climate finance, banks and insurance industry, are to name a few]. In summary, the enterprise of building human resources in weather and climate modelling has not only enormous benefits both for advancing scientific research in atmospheric / oceanic sciences, but also for the overall progress of the society and nation at large.